\documentclass[aps,prd,twocolumn,showpacs,groupedaddress,nofootinbib,letterpaper]{revtex4}

\usepackage{amsmath,amsthm,amssymb,enumerate}
\usepackage{graphicx}
\usepackage{hyperref}
\usepackage{bbold}

\newcommand{\avg}[1]{\left\langle#1\right\rangle}

\newcommand{\ket}[1]{\left\vert#1\right\rangle}
\newcommand{\braket}[2]{\left\langle#1\middle|#2\right\rangle}
\newcommand{\ketbra}[2]{\left|#1\middle\rangle \! \middle\langle#2\right|}

\newcommand{\oneloop}{\square}
\newcommand{\twoloop}{\square\!\square}
\newcommand{\wideloop}{\sqsubset\!\sqsupset}

\begin{document}

\title{Decomposition of entanglement entropy in lattice gauge theory}
\author{William Donnelly}
\affiliation{
Center for Fundamental Physics, Department of Physics, \\
University of Maryland, College Park, Maryland 20742-4111, USA
}
\email{wdonnell@umd.edu}

\begin{abstract}
We consider entanglement entropy between regions of space in lattice gauge theory.
The Hilbert space corresponding to a region of space includes edge states that transform nontrivially under gauge transformations.
By decomposing the edge states in irreducible representations of the gauge group, the entropy of an arbitrary state is expressed as the sum of three positive terms: a term associated with the classical Shannon entropy of the distribution of boundary representations,
a term that appears only for non-Abelian gauge theories and depends on the dimension of the boundary representations, 
and a term representing nonlocal correlations.
The first two terms are the entropy of the edge states, and depend only on observables measurable at the boundary.
These results are applied to several examples of lattice gauge theory states, including the ground state in the strong coupling expansion of Kogut and Susskind.
In all these examples we find that the entropy of the edge states is the dominant contribution to the entanglement entropy.
\end{abstract}


\maketitle

\section{Introduction}

In quantum field theory, regions of space are subsystems, and the entropy of these subsystems, which for pure states is the entanglement entropy, gives important information about the quantum state.
The entanglement entropy plays an important role in black hole physics, where it governs the one-loop corrections to the Bekenstein-Hawking entropy, and may in fact be responsible for the entire black hole entropy \cite{Bombelli1986,Srednicki1993,Susskind1994,Jacobson1994}.

In addition to its role in black hole statistical mechanics, entanglement entropy has been found to play a role in the study of phase transitions.
Entanglement entropy can be used to test for the presence of long-range order, even when an order parameter is not known \cite{Osborne2002a,Vidal2003} and can detect the presence of topological phases \cite{Kitaev2005, Levin2006}.
In gauge theories, entanglement entropy may be useful in studying the deconfining phase transition \cite{Nishioka2006, Klebanov2007}.

In gauge theories, the entanglement entropy is complicated by the fact that states are not precisely localizable in space.
The result is the the Hilbert space corresponding to a region of space includes edge states that contribute to the entanglement entropy \cite{Balachandran1994, Buividovich2008b}.
These edge states are similar to the ``would-be pure gauge'' degrees of freedom in (2+1)-dimensional quantum gravity \cite{Carlip1994}.
Recall that in 2+1 gravity there are no local degrees of freedom in the bulk, yet the usual thermodynamic arguments suggest that black holes have an entropy proportional to the length of the horizon. 
In the approach of Ref.~\cite{Carlip1994}, the horizon has local degrees of freedom and it is these degrees of freedom whose entropy is given by the Bekenstein-Hawking area law.
While the precise relation between these two notions of boundary states is not clear, their similarity suggests a relation between the entanglement entropy and the boundary state counting method of Ref.~\cite{Carlip1994}.

It is well-known that the entanglement entropy is ultraviolet divergent, so a regulator is needed in its definition.
In the continuum, standard methods for calculating the entanglement entropy at one-loop order fail for gauge fields \cite{Kabat1995}.
The Hamiltonian method does not work because the eigenfunctions of the boost generator in Rindler space are not square-integrable. 
The Euclidean conical deficit angle method leads to a ``contact term'' that makes the result negative and therefore not identifiable with entanglement entropy, which is a manifestly positive quantity.
In this paper we consider the entanglement entropy for lattice gauge theories.
This allows us to regulate the entanglement entropy while clearly exposing the role of gauge invariance.

Entanglement entropy has been considered before for certain classes of states in lattice gauge theory \cite{Buividovich2008a, Buividovich2008b, Buividovich2008c, Velytsky2008a, Velytsky2008b}.
This paper adopts the Hamiltonian formulation of lattice gauge theory, rather than the replica method \cite{Callan1994} that has typically been used in numerical calculations of the entanglement entropy.
The replica method relates the entanglement entropy of the ground state of a given theory to the partition function computed on an $n$-sheeted cover of the Euclidean spacetime.
Our results do not use the replica method, so they do not require the state to be expressed as a Euclidean path integral, though we agree with results obtained using the replica method where the latter is applicable.

Closely related to lattice gauge theory is loop quantum gravity, which is formulated as an SU(2) lattice gauge theory on a superposition of lattices.
Although this paper will not discuss loop quantum gravity, entanglement entropy in loop quantum gravity was discussed in Refs.~\cite{Dasgupta2005,Donnelly2008}, and we expect the techniques of this paper to generalize easily to a superposition of lattices.
We note also that the Hilbert space of edge states in SU(2) lattice gauge theory is closely related to the Hilbert space of the SU(2) Chern-Simons theory whose states are counted in the loop quantum gravity derivation of black hole entropy \cite{Ashtekar1997,Engle2009}.

We now briefly summarize our result.
Consider a lattice whose set of nodes $N$ is divided into two disjoint sets $A$ and $B$ whose union is all of $N$.
In a lattice theory where the degrees of freedom live on the nodes, the Hilbert space associated with a set of nodes is simply the tensor product of the Hilbert spaces of each individual node. 
This leads to a tensor product decomposition of the whole Hilbert space as ${\cal H} = {\cal H}_A \otimes {\cal H}_B$.
In a lattice gauge theory, the degrees of freedom live on the links, so there is not such a simple tensor product decomposition.
However, following Ref.~\cite{Buividovich2008b} we can define a Hilbert space ${\cal H}_A$ by splitting the links that cross the boundary.
Along each link $l$ with one end point in $A$ and one end point in $B$, we insert a new vertex on the boundary and divide the link into two smaller links, one associated with region $A$ and one associated with region $B$.
The Hilbert space ${\cal H}_A$ then consists of functionals of the connection on the links in $A$ that are invariant under gauge transformations that act on the nodes in the interior of $A$, but not on the boundary nodes.
In restricting the gauge symmetry, degrees of freedom that were previously pure gauge are promoted to physical degrees of freedom.
The new degrees of freedom are edge states that are associated with the boundary vertices and transform nontrivially under gauge transformations acting on the boundary.
They are the lattice analogue of the continuum edge states studied in Ref.~\cite{Balachandran1994}.

The Hilbert space $\cal H$ is not equal to ${\cal H}_A \otimes {\cal H}_B$, since the former is invariant under all gauge transformations, and the latter is invariant under only those gauge transformations that act trivially on the boundary.
Thus instead of an isomorphism of Hilbert spaces, we have the embedding
\begin{equation} \label{embedding}
{\cal H} \to {\cal H}_A \otimes {\cal H}_B.
\end{equation}
The entanglement entropy of any state in $\cal H$ can be defined by embedding the state into ${\cal H}_A \otimes {\cal H}_B$.
Letting $\rho$ denote a state, represented as a density matrix in ${\cal H}_A \otimes {\cal H}_B$, the reduced density matrix of system $A$ is the partial trace $\rho_A = \text{tr}_B (\rho)$, and the entanglement entropy is
\begin{equation}
S_\text{entanglement} = S(\rho_A)
\end{equation}  
where the function $S(\rho)$ is the von Neumann entropy,
\begin{equation}
S(\rho) = - \text{tr} (\rho \ln \rho).
\end{equation}

The states in $\cal H$ are invariant under all gauge transformations, including those acting on the boundary.
The reduced density matrix $\rho_A$ associated with a gauge-invariant state is then also invariant under the group of boundary gauge transformations, which acts nontrivially on ${\cal H}_A$.
When decomposed into irreducible representations of the group of boundary gauge transformations, the matrix $\rho_A$ takes the form of a direct sum of tensor products.
Using properties of the von Neumann entropy under direct sum and tensor product, we decompose the entanglement entropy of a generic state in lattice gauge theory as a sum of three positive terms \eqref{threeterms}:
\begin{itemize}
\item the Shannon entropy of the distribution of boundary link representations found in Ref.~\cite{Buividovich2008b}, 
\item the weighted average of the logarithm of the dimension of the boundary representations found in Ref.~\cite{Donnelly2008}, and 
\item a third term that captures nonlocal correlations of the bulk field.
\end{itemize}
The first two terms are purely local to the boundary, and together they capture the entropy of the edge states.
Our result directly generalizes the results of Refs.~\cite{Buividovich2008b,Donnelly2008}.
We will derive the result in Sec.~\ref{section:main} and give several applications to specific states of lattice gauge theory in Sec.~\ref{section:examples}.

\section{Entanglement entropy in lattice gauge theory} \label{section:main}

We first review the kinematics of Hamiltonian lattice gauge theory \cite{Kogut1975} and spin network states \cite{Baez1994}.
Consider a lattice consisting of a set $N$ of nodes and a set $L$ of oriented links, and let $G$ be a gauge group that is either a compact Lie group, or a discrete group.
A field configuration is an assignment of group elements $u_l$ to links, and a gauge transformation is an assignment of group elements $g_n$ to nodes, which acts on $u_l$ as
\begin{equation}
u_l \to g_{t(l)}^{\phantom{-1}} \circ u_l \circ g_{s(l)}^{-1}
\end{equation}
where $s(l)$ and $t(l)$ are respectively the nodes at the source and target of the link $l$.
The Hilbert space $\cal H$ consists of square-integrable functionals of the holonomies $u_l$ that are invariant under gauge transformations.
An orthonormal basis for $\cal H$ is given by a generalization of the spin network states \cite{Penrose1971, Rovelli1995, Baez1994}.
A spin network consists of an assignment of irreducible representations $R = \{ r_l : l \in L\}$ to each link, and intertwiners $I = \{i_n : n \in N\}$ to each node.
Each intertwiner $i_n$ is a $G$-invariant linear map between representation spaces
\begin{equation}
i_n : \left( \bigotimes_{l : t(l) = n} r_l \right) \to \left( \bigotimes_{l : s(l) = n} r_l  \right)
\end{equation}
The spin network state $\ket{S}$ associated with a spin network $S$ is the functional obtained by taking the representation $r_l$ of the group element on each link $l$, multiplying by $\sqrt{\dim(r)}$, and contracting the free indices with the intertwiners $i_n$ \cite{Baez1994},
\begin{equation} \label{wavefunction}
\braket{S}{U} = \left( \bigotimes_{l \in L} \sqrt{\dim(r_l)} \, r_l(u_l) \right) \circ \left( \bigotimes_{n \in N} i_n \right).
\end{equation}
The intertwiners are chosen to be orthonormal in the inner product,
\begin{equation}
\langle i_1, i_2 \rangle = \text{tr}(i_1 i_2^\dag)
\end{equation}
so that the resulting spin network states form an orthonormal basis of $\cal H$ \cite{DePietri1996}.

We now describe the tensor product decomposition of the Hilbert space, which was described in Ref.~\cite{Buividovich2008b} for Abelian lattice gauge theories.
Let $A$ be a region of space, which on the lattice will mean a subset of the nodes.
The configuration space of $\cal H$ consists of holonomies on all links of the lattice.
The links can be divided into three sets: $L_A$ is the set of links with both end points in $A$, $L_B$ is the set of links with both end points in $B$ and $L_{\partial}$ is the set of links that cross the boundary.                                                                                                                                                                               
In order to partition the degrees of freedom of the boundary links between ${\cal H}_A$ and ${\cal H}_B$, we split each boundary link into two at the boundary, such for each link in $L_{\partial}$ there is a new link in $L_{\partial A}$ and one in $L_{\partial B}$.
The Hilbert space ${\cal H}_A$ is then defined as the square-integrable functions of the holonomies $\{ u_l : l \in L_A \cup L_{\partial A} \}$ invariant under gauge transformations acting at nodes in the interior of $A$ (but \emph{not} under gauge transformations acting on the boundary).
Then for each link $l \in L_{\partial}$, the holonomy $u_l$ can be obtained as a product of a holonomy in $L_{\partial A}$ and one in $L_{\partial B}$, and we define the product map
\begin{equation}
\pi : U_{\partial A} \times U_{\partial B} \to U_{\partial}.
\end{equation}
The pullback map $\pi^*$ then gives an embedding
\begin{equation}
\pi^* : \cal H \to {\cal H}_A \otimes {\cal H}_B.
\end{equation}
For example, if there is just one boundary link $l$ split into links $l_1$ and $l_2$ such that $t(l_1) = s(l_2)$, then the map $\pi$ is given by
\begin{equation}
\pi(u_1, u_2) = u_2 u_1,
\end{equation}
and the pullback $\pi^*\psi$ of a function $\psi: G \to \mathbb{C}$ is given by
\begin{equation}
(\pi^* \psi) (u_1, u_2) = \psi(u_1 u_2).
\end{equation}
This embedding preserves the norm of the state, a fact which follows from the $G$-invariance and normalization of the Haar measure.

To specify a spin network $S$, we specify all its representations and intertwiners
\begin{equation} \label{S}
S = (R_A, R_B, R_{\partial}, I_A, I_B).
\end{equation}
Just as the space $\cal H$ is spanned by spin network states, the space ${\cal H}_A$ is spanned by open spin network states \cite{Ashtekar1996}.
An open spin network $S_A$ is specified by
\begin{equation} \label{SA}
S_A = (R_A, R_{\partial}, I_A, M)
\end{equation}
where $M = \{m_l : l \in L_{\partial}\}$ is a set of vectors in the boundary representation spaces, $m_l \in r_l$ if the link $l$ points inward at the boundary, or in the dual representation $m_l \in \bar{r}_l$ if the link $l$ points outward (for unitary representations, the dual representation $\bar{r}$ and complex conjugate representation $r^*$ coincide).
The open spin network state $\ket{S_A}$ is defined just as in Eq.~\eqref{wavefunction}, except that the extra free indices associated with the boundary vertices are contracted with the vectors $m_l$.
The open spin network states form an orthonormal basis of ${\cal H}_A$ provided the vectors $m_l$ and intertwiners $i_n$ are chosen to be orthonormal.

As shown in Ref.~\cite{Donnelly2008}, under the embedding $\pi^*$, the spin network state $\ket{S}$ maps to
\begin{equation} \label{schmidt}
\pi^* \ket{S} =
\prod_{l \in L_{\partial A}} \frac{1}{\sqrt{\dim(r_l)}} \sum_{m_l}
\ket{S_A} \otimes \ket{S_B}
\end{equation}
where $S_A$ is given by \eqref{SA}, and $S_B = (R_B, R_{\partial}, I_B, M^*)$, where the vectors $M^*$ are dual (complex conjugate) to the vectors $M$.
In Eq.~\eqref{schmidt}, $m_l$ ranges over an orthonormal basis of $r_l$.
This decomposition follows from inserting a resolution of unity at each point where a link crosses the boundary, and the factors of $1/\sqrt{\dim(r_l)}$ arise to cancel the extra factors of $\sqrt{\dim(r_l)}$ in Eq.~\eqref{wavefunction} that come from splitting the boundary links.

We now consider an arbitrary gauge-invariant state $\ket{\psi}$ expressed in the spin network basis,
\begin{equation}
\ket{\psi} = \sum_S \psi(S) \ket{S}.
\end{equation}
Using the decomposition \eqref{schmidt}, the reduced density matrix for region $A$ is
\begin{equation} \label{rhoa}
\rho_A = \sum_{\stackrel{R_A,R_A',I_A, I_A', }{R_{\partial},R_B, I_B, M}} 
\frac
{\psi(S) \psi(S')^*}
{\prod_{l \in L_{\partial A}} \text{dim}(r_l) } 
\ketbra{S_A}{S'_A}
\end{equation}
where $S$ and $S_A$ are given by \eqref{S} and \eqref{SA}, and $S'$ and $S'_A$ are given by
\begin{align}
S' &= (R'_A, R_B, R_{\partial}, I'_A, I_B), & S'_A &= (R'_A, R_{\partial}, I'_A, M).
\end{align}
The sums over intertwiners in Eq.~\eqref{rhoa} are taken over an orthonormal basis of the space of intertwiners compatible with the representations incident on each node. 
In the case where there is no such intertwiner, the sum is zero.

There are two features worthy of note about Eq.~\eqref{rhoa}.
First, the set of representations $R_{\partial}$ is always the same for $S_A$ and $S'_A$.
This means that the matrix $\rho_A$ has no off-diagonal terms that mix different boundary representations.
Second, the coefficients in Eq.~\eqref{rhoa} are independent of $M$, so within each representation the $M$ degrees of freedom are in a maximally mixed state.

This structure of the reduced density matrix can be seen from group theory.
The boundary gauge transformations form a group $G^n$ where $n$ is the number of boundary links, and is represented unitarily on ${\cal H}_A$.
Such a representation $R(g)$ can always be written as a direct sum of irreducible representations:
\begin{equation} \label{u}
R(g) = \bigoplus_r r(g) \otimes \mathbb{1}_{n(r)}
\end{equation}
where $g \in G^n$, $r$ runs over all irreducible representations of $G^n$, and $n(r)$ is the multiplicity with which the irreducible representation $r$ appears in the representation $R$.
The reduced density matrix $\rho_A$ comes from a gauge-invariant state, so it must commute with the representation $R$.
To commute with $R(g)$ for all $g$, $\rho_A$ must take the form
\begin{equation}
\rho_A = \bigoplus_r \frac{\mathbb{1}_{\dim(r)}}{\dim(r)} \otimes \rho_A(r)
\end{equation}
where $\rho_A(r)$ is a density matrix of dimension $n(r)$.

To see more explicitly how the density matrix decomposes into representations, it is useful to divide the Hilbert space ${\cal H}_A$ into an edge Hilbert space and a bulk Hilbert space such that the states of the boundary Hilbert space are labeled by $(R_{\partial}, M)$, and the bulk Hilbert space is labeled by $(R_A, I_A)$.
This decomposition of the Hilbert space is
\begin{equation} \label{decomp}
{\cal H}_A = \bigoplus_{R_{\partial}} \left[ 
\left( \bigotimes_{l \in L_{\partial}} r_l \right) \otimes
{\cal H}_A(R_{\partial})
\right]
\end{equation}
where ${\cal H}_A(R_{\partial})$ is spanned by states $\ket{R_A, I_A}$ that are compatible with the assignment of representations $R_{\partial}$ to the boundary.

In the decomposition \eqref{decomp} the open spin network states $\ket{S_A}$ and $\ket{S'_A}$ can be written as
\begin{eqnarray}
\ket{S_A} &= \ket{R_{\partial}} \otimes \ket{M} \otimes \ket{R_A,I_A}, \\
\ket{S_A'} &= \ket{R_{\partial}} \otimes \ket{M} \otimes \ket{R_A',I_A'}.
\end{eqnarray}
so that their outer product takes the form
\begin{equation} \label{outer}
\ketbra{S_A}{S_A'} = \ketbra{R_{\partial}}{R_{\partial}} \otimes \ketbra{M}{M} \otimes \ketbra{R_A,I_A}{R_A',I_A'}.
\end{equation}
Substituting Eq.~\eqref{outer} into the reduced density matrix \eqref{rhoa} and rearranging terms yields
\begin{equation} \label{rhoa2}
\rho_A = \sum_{R_{\partial}} p(R_{\partial}) \ketbra{R_{\partial}} {R_{\partial}} 
\otimes \left( \sum_M \frac{\ketbra{M}{M}}{\prod_{l \in L_{\partial A}} \text{dim}(r_l) } \right) 
\otimes \rho_A(R_\partial)
\end{equation}
where $p(R_{\partial})$ is the probability of distribution of representations on the boundary,
\begin{equation}
p(R_{\partial}) = \sum_{R_A,R_B, I_A, I_B} \left| \psi(S) \right|^2,
\end{equation}
and $\rho_A(R_{\partial})$ is the reduced density matrix,
\begin{equation}
\rho_A(R_{\partial}) = 
\sum_{\stackrel{R_A, R_A', R_B,}{I_A, I_A', I_B}} 
\frac{\psi(S) \psi(S')^*}{p(R_{\partial})}
\ketbra{R_A, I_A}{R'_A, I'_A}.
\end{equation}
The factor $p(R_{\partial})$ is included in the definition of $\rho_A(R_{\partial})$ to maintain the unit trace condition.

Since the same $R_{\partial}$ appears in both the ket and the bra in the first tensor factor in Eq.~\eqref{rhoa2}, the state does indeed lie in the direct sum Hilbert space \eqref{decomp}.
Moreover, the second tensor factor in Eq.~\eqref{rhoa2} is proportional to the identity matrix, so the density matrix $\rho_A$ can equivalently be written
\begin{equation} \label{directsum}
\rho_A = \bigoplus_{R_{\partial}} p(R_{\partial}) \left[ 
\left( \bigotimes_{l \in L_{\partial}} \frac{\mathbb{1}_{r_l}}{\dim(r_l)} \right)
\otimes \rho_A(R_{\partial})
 \right].
\end{equation}

The structure of the reduced density matrix \eqref{directsum} allows us to simplify the entanglement entropy by using properties of the von Neumann entropy under direct sum and direct product.
Let $p_n$ be positive real numbers summing to one, and $\rho_n$ density matrices on Hilbert space ${\cal H}_n$.
The von Neumann entropy of a weighted direct sum is
\begin{equation} \label{ssum}
S \left( \bigoplus_n p_n \rho_n \right) = H(p_n) + \avg{ S(\rho_n) }
\end{equation}
where $\avg{\cdot}$ denotes expectation value with respect to the probability distribution $p_n$, and $H(p_n)$ is the Shannon entropy of this distribution (the classical analogue of the von Neumann entropy),
\begin{equation}
H(p_n) = -\sum_n p_n \ln p_n.
\end{equation}
Under a tensor product, the von Neumann entropy is additive,
\begin{equation} \label{sproduct}
S(\rho_1 \otimes \rho_2) = S(\rho_1) + S(\rho_2).
\end{equation}
Finally, the maximally mixed state of dimension $n$ has entropy $\ln n$,
\begin{equation} \label{smax}
S( \mathbb{1}_n / n) = \ln n.
\end{equation}

Applying the properties of the von Neumann entropy \eqref{ssum}, \eqref{sproduct} and \eqref{smax} to the reduced density matrix $\rho_A$ \eqref{directsum} gives the entropy as the sum of three positive terms,
\begin{equation} \label{threeterms}
S(\rho_A) = H(p(R_{\partial})) + \sum_{l \in L_{\partial A}} \!\!\! \avg{\ln \text{dim}(r_l)} + \avg{S(\rho_A(R_{\partial}))}
\end{equation}
where $\avg{\cdot}$ denotes expectation value with respect to the probability distribution $p(R_{\partial})$.
Eq.~\eqref{threeterms} is the main result of this paper.
Individual terms in this expression have appeared before: the first term appeared in Ref.~\cite{Buividovich2008b} where it was derived for a specific class of states (see section \ref{section:electricstrings}), and the second term appeared in Ref.~\cite{Donnelly2008} as the entropy of a single spin network state.

The first two terms of Eq.~\eqref{threeterms} depend only on the distribution of the boundary representations, and in this sense are purely local.
The second term is a sum over boundary links, and so is extensive on the boundary.
The first term is not extensive, but will be approximately extensive as long as the correlations between different representations are local.
The effect of correlations is always to decrease the entropy, so we can obtain an extensive upper bound by neglecting these correlations.
If we assume that the statistics of each edge are the same (which would be the case for states with a discrete translation and rotation symmetry, such as the ground state of a translation- and rotation-invariant Hamiltonian) then the upper bound depends only on the probability distribution of representation on each edge $p(r)$ and is given by
\begin{equation}
S_\text{boundary} \leq n \Big( H(p) + \avg{\ln \dim (r)} \Big),
\end{equation}
with $n$ the number of boundary links.

In principle either of the local terms can be larger.
For example, in a state sharply peaked on spin networks with high-dimension representations, the second term will dominate. 
In a state that is a superposition of many spin networks with low-dimension representations, the first term will dominate.
In particular, an Abelian theory has only one-dimensional representations so the second term in Eq.~\eqref{threeterms} vanishes.

The third term in Eq.~\eqref{threeterms} is the most difficult to characterize.
It includes the effects of correlations between distantly separated degrees of freedom, and in general it is not bounded by the area of the boundary.
However we will see that, for the classes of states considered in Sec.~\ref{section:examples}, this term is either vanishing or much smaller than the local terms.

\section{Examples} \label{section:examples}
We now consider several examples of states whose entanglement entropy can be calculated using our method.

\subsection{Electric string states} \label{section:electricstrings}

Reference \cite{Buividovich2008b} considers a class of states in ${\mathbb Z}_2$ lattice gauge theory.
This theory has just two irreducible representations: a trivial representation and an alternating representation.
The associated spin network states are ``electric string states'' where the two representations are interpreted as the presence or absence of electric strings along the edges.
The states considered are of the form
\begin{equation} \label{electricstring}
\ket{\alpha} = \frac{1}{\cal N}\sum_S e^{-\frac{\alpha}{2} L(S)} \ket{S}
\end{equation}
where $\alpha$ is a real parameter, $L(S)$ is the total length of electric string (i.e. the number of alternating representations), and $\cal N$ is a normalization factor.

For such a state, we now show that the entanglement entropy is given entirely by the Shannon entropy of the representations on the boundary [the first term in Eq.~\eqref{threeterms}].
Since the gauge group is Abelian, the second term in Eq.~\eqref{threeterms} vanishes.
Now consider fixing the set of representations on the boundary, $R_{\partial}$.
The total length of electric strings is the sum of the strings in $A$, the strings in $B$, and those crossing the boundary: $L(S) = L(S_A) + L(S_B) + L(\partial A)$.
For a fixed set of boundary representations, the reduced density matrix is
\begin{eqnarray}
\rho_A(R_{\partial}) &\propto& \sum_{S_A, S'_A} e^{-\frac{\alpha}{2} L(S_A) -\frac{\alpha}{2} L(S'_A)} \ketbra{S_A}{S'_A} \\
&=& \ketbra{\psi}{\psi}
\end{eqnarray}
where 
\begin{equation}
\ket{\psi} = \sum_{S_A} e^{-\frac{\alpha}{2} L(S_A)} \ket{S_A}.
\end{equation}
This is a pure state, so $S(\rho_A(R_{\partial})) = 0$.

Here we rederive the result that for the states in Eq.~\eqref{electricstring} the entropy is just the Shannon entropy of the string end points.
The entropy for this special class of states was originally derived in Ref.~\cite{Buividovich2008b}, and on that basis it was conjectured that the Shannon entropy of the string end points is a good approximation to the full entanglement entropy of the ground state in lattice gauge theory.
The advantage of the Shannon entropy over the full entropy is that it depends only on the probability distribution of representations on the boundary, and is more easily computed in computer simulations.

Here we have proven that the Shannon entropy is a lower bound to the entropy that appears generically for all states, and not just states of the special form \eqref{electricstring}.
Moreover we have characterized precisely the difference between the Shannon entropy and the full entanglement entropy.
The fact that Ref.~\cite{Buividovich2008b} finds good agreement between the Shannon entropy and the full entropy for the true ground state indicates that the third term in Eq.~\eqref{threeterms} is subleading for this state.

For a non-Abelian gauge theory we can improve on the Shannon entropy as an approximation of the full entropy by including the log-dimension term [the second term of Eq.~\eqref{threeterms}].
This term also depends only on the distribution of boundary representations and should therefore also be easy to compute in computer simulations.
Reference \cite{Buividovich2008b} also noted the similarity of the Shannon entropy to the log-dimension term that appeared in Ref.~\cite{Donnelly2008}.
Here we have shown that these two terms are distinct contributions to the entanglement entropy.

\subsection{Topological phase ground state}

In Ref. \cite{Levin2006}, the limit $\alpha \to 0$ was considered, in which the state approaches a superposition of spin network states in which every spin network has an equal amplitude.
Every configuration of string end points on the boundary of a region is equally probable, but gauge invariance requires the total number of string end points crossing each connected component to be even.
For a region with $n$ boundary edges and whose boundary has $k$ connected components, the entropy is
\begin{equation} \label{alphazero}
S = (n - k) \ln 2.
\end{equation}
For a macroscopic region, $n$ becomes large while $k$ stays constant so the entropy is approximately extensive on the boundary, with small nonextensive corrections.

The deviation from extensivity of the entanglement entropy is captured by the topological entanglement entropy.
Given a pair of regions $A$ and $B$,\footnote{We are now allowing $B$ to be an arbitrary region, not necessarily the complement of $A$.} the topological entanglement entropy is (following Ref.~\cite{Levin2006}, but closely related to the definition in Ref.~\cite{Kitaev2005})
\begin{equation}
S_\text{top} = S(A) + S(B) - S(A \cup B) - S(A \cap B).
\end{equation}
Note that terms proportional to the volume or to the surface area [such as the term proportional to $n$ in Eq.~\eqref{alphazero}] do not contribute to the topological entanglement entropy.
This is because the volume and surface area obey the inclusion-exclusion principle,
\begin{equation}
f(A) + f(B) = f(A \cup B) + f(A \cap B).
\end{equation}
where $f$ is a function measuring either the volume or the surface area\footnote{Note that the Euler characteristic also obeys the inclusion-exclusion principle.
This means that (counterintuitively) terms in the entanglement entropy proportional to the Euler characteristic do not contribute to the topological entanglement entropy.}.
However the $k$-dependent term does contribute to the topological entanglement entropy.
If one considers a set of regions $A$, $B$ as in Ref.~\cite{Levin2006} such that $A$ and $B$ are each topologically disks, $A \cap B$ has two connected components, and $A \cup B$ is topologically an annulus, we find a topological entanglement entropy of
\begin{equation}
S_\text{top} = 2 \ln 2
\end{equation}
in agreement with the result of Ref.~\cite{Levin2006}.

\subsection{Strong coupling limit}

We now consider the entanglement entropy of the ground state of the $\text{SU}(2)$ Kogut-Susskind Hamiltonian \cite{Kogut1975} in the limit of strong coupling, $g \gg 1$.
Consider a hypercubic lattice in dimension $d \geq 2$.
The Kogut-Susskind Hamiltonian is a sum of electric and magnetic parts,
\begin{equation}
H = H_E + H_B.
\end{equation}
We will work with a rescaled version of this Hamiltonian, but the ground state and therefore its entanglement entropy are not sensitive to this rescaling.
The electric part is diagonal in the spin network basis, and is given by
\begin{equation}
H_E \ket{S} = 
\sum_{l \in L} j_l (j_l + 1) \ket{S}
\end{equation}
where $j_l$ is the spin of the representation $r_l$.
The state of lowest energy for $H_E$ is the spin network state in which all edges are in the $j=0$ representation.
We will denote this state by $\ket{0}$.

The magnetic part of the Hamiltonian is not diagonal in the spin network basis, but can be expressed as a functional of the holonomies,
\begin{equation}
H_B = 
3 \lambda \sum_{\oneloop} \left[ \text{tr}(u_\oneloop) + \text{h.c.} \right]
\end{equation}
where $\square$ is the set of all plaquettes (closed loops containing exactly four links), and $\text{tr}(u_\square)$ is the associated Wilson loop operator in the fundamental representation $j = \tfrac{1}{2}$. 
The parameter $\lambda$ is related to the gauge coupling $g$ by $\lambda \sim g^{-4}$.
The operator $\text{tr}(u_\square)$ acts on the trivial spin network as
\begin{equation}
\text{tr}(u_\square) \ket{0} = \ket{\square}
\end{equation}
where $\ket{\square}$ is the spin network state in which each edge around the plaquette $\square$ is assigned the $j= \tfrac{1}{2}$ representation, and all other edges are assigned the trivial representation.

For strong coupling $g \gg 1$, so $\lambda \ll 1$ and we can use perturbation theory to calculate the ground state, treating $H_B$ as a perturbation of $H_E$. 
We will be interested in computing the entropy to order $\lambda^2$, so we compute the ground state to order $\lambda^2$:
\begin{eqnarray} \label{groundstate}
\ket{\Omega} &=& 
\left(1 - \tfrac{1}{2} N_\oneloop \lambda^2 \right) \ket{0} + \lambda \sum_{\oneloop} \ket{\oneloop} 
\nonumber \\
&& + \lambda^2 \left(
\sum_{\oneloop, \oneloop'} \ket{\oneloop \oneloop'} +
\sum_{\twoloop} c_{\twoloop} \ket{\twoloop} + \sum_{\wideloop} c_{\wideloop} \ket{\wideloop} \right) \nonumber \\
&& + O(\lambda^3).
\end{eqnarray}
Here $N_\oneloop$ is the total number of plaquettes in the lattice, ensuring that the state is normalized to order $\lambda^2$.
The state $\ket{\oneloop \oneloop'}$ denotes the spin network state of two nonintersecting single-plaquette Wilson loops around the plaquettes $\oneloop$ and $\oneloop'$.
The state $\ket{\twoloop}$ denotes a spin network state with support on two intersecting plaquettes with outer links in the $j=\tfrac{1}{2}$ and an intermediate link with $j=1$, and the state $\ket{\wideloop}$ is a spin network of a single loop encircling two plaquettes in the $j=\tfrac{1}{2}$ representation.
The numbers $c_{\twoloop}$ and $c_{\wideloop}$ are constants of order unity that are irrelevant for the entanglement entropy.

To describe the way different spin networks intersect the region $A$, we will write $\oneloop \in A$, $\twoloop \in A$ to indicate spin networks that lie entirely in region $A$.
We can divide the single-plaquette spin networks into those within $A$, those within $B$, and those intersecting the boundary.
The numbers of plaquettes of each type are given by $N_\oneloop(A)$, $N_\oneloop(B)$, and $N_\oneloop(\partial)$, respectively, with
\begin{equation}
N_\oneloop = N_\oneloop(A) + N_\oneloop(B) + N_\oneloop(\partial).
\end{equation}
To calculate $N_\oneloop(\partial)$, we note that a single-plaquette loop, if it intersects the boundary at all must intersect an even number of times. 
We will assume that the region $A$ is chosen so that single-plaquette loops can intersect either twice or not at all.
Let $n$ be the number of boundary links of region $A$.
To count the number of ways a single plaquette can intersect the boundary, we fix one of the links intersecting the boundary, and after doing so there are $2 (d-1)$ different orientations the plaquette can take.
This overcounts by a factor of $2$, since the loop intersects the boundary twice, and so there are
\begin{equation}
N_\oneloop(\partial) = n(d-1)
\end{equation}
ways a single plaquette can intersect the boundary.

We now compute the entanglement entropy of the state \eqref{groundstate} by calculating each term of Eq.~\eqref{threeterms} in turn.
To find the probability distribution of representations on the boundary we note that the probability of a two-plaquette state is $O(\lambda^4)$ and therefore negligible.
Thus the only states contributing to this distribution are the trivial spin network and the single-plaquette spin networks.
The number of different possible sets of representations $R_\partial$ is $N_\oneloop(\partial)$ and each has probability $\lambda^2$, with the probability of having no intersections given by $1 - N_\oneloop(\partial) \lambda^2$.
The entropy of this probability distribution is
\begin{eqnarray}
H(p(R_\partial))
&=& n (d-1) \lambda^2 ( -\ln \lambda^2 + 1) + O(\lambda^3)
\end{eqnarray}
and since each single-plaquette spin network intersects in two $j=\tfrac{1}{2}$ links, the second term of Eq.~\ref{threeterms} is
\begin{equation}
\sum_{l \in L_{\partial}} \avg{\ln (2 j_l + 1)} = n (d-1) \lambda^2 2 \ln 2.
\end{equation}

We now consider the entropy of the density matrices $\rho_A(R_{\partial})$.
Since we are taking an expectation value, we only need to consider sets of boundary representations with probability of order $\lambda^2$ or larger.
This means either there is no intersection with the boundary, or a single plaquette $\oneloop$ intersecting the boundary.
In the latter case, the only matrix element of $\rho_A$ compatible with the assignment of representations to the boundary and probability at least order $\lambda^2$ is $\lambda^2 \ketbra{\oneloop}{\oneloop}$.
Thus $\rho_A(R_{\partial})$ is a pure state to order $\lambda^2$ and so contributes no entropy.

In the case where there is no plaquette intersecting the boundary, we need to know the state $\rho_A(R_{\partial})$ to order $\lambda^2$.
A short calculation shows that 
\begin{equation}
\rho_A(R_{\partial}) = \ketbra{\psi}{\psi} + O(\lambda^3)
\end{equation}
where
\begin{eqnarray}
\ket{\psi} &=& (1 - \tfrac{1}{2} N_\oneloop(A) \lambda^2 ) \ket{0} + \lambda \sum_{\oneloop \in A} \ket{\oneloop} \\
&& + \lambda^2 \left(
\sum_{\oneloop, \oneloop' \in A} \!\!\! \ket{\oneloop \oneloop'} +
\!\!\! \sum_{\twoloop \in A} \!\!\! c_{\twoloop} \ket{\twoloop} +
\!\!\! \sum_{\wideloop \in A} \!\!\! c_{\wideloop} \ket{\wideloop} \right). \nonumber
\end{eqnarray}
Since this is a pure state, its entropy is zero to order $\lambda^2$.

Combining the terms in the previous paragraphs the entanglement entropy at first nonvanishing order in the strong coupling expansion is
\begin{equation}
S = n (d-1) \lambda^2 ( -\ln \lambda^2 + 1 + 2 \ln 2) + O(\lambda^3).
\end{equation}
It is extensive in the boundary area (proportional to $n$).
The entropy is also proportional to $(d-1)$, which is the number of polarizations of the gauge field.
This factor is to be expected for weak coupling, where free field theory is a good approximation.
It is not clear why this factor should appear also at strong coupling.

\section{Conclusion}

We have given a formula for the entanglement entropy of an arbitrary state in lattice gauge theory as a sum of three terms [Eq.~\eqref{threeterms}].
Two of these terms are local to the boundary and have appeared before in the literature \cite{Buividovich2008b, Donnelly2008}; the other captures nonlocal correlations between bulk degrees of freedom.
Our result extends the result of Ref.~\cite{Buividovich2008b}, which proposed that the Shannon entropy of the boundary representations [the first term of Eq.~\eqref{threeterms}] is an approximation to the entanglement entropy that depends only on the statistics of boundary observables.
Our results prove that the Shannon entropy is a lower bound, and we give an improvement of this lower bound for non-Abelian gauge theories [the second term of Eq.~\eqref{threeterms}] that also depends only on the statistics of boundary observables.
Moreover, a precise expression is given for the difference between the local part of the entropy and the full entropy [the third term of Eq.~\eqref{threeterms}].

We have verified several results for entanglement entropy of specific states that appeared already in the literature, and considered also the entanglement entropy of the ground state of the Kogut-Susskind Hamiltonian for SU(2) lattice gauge theory to first nonvanishing order in the strong coupling expansion.
While at this leading order only the local terms contribute to the entropy, at higher order all terms will contribute.
This agrees with field theory calculations of the entropy, where the entropy density is found to diverge as the horizon is approached.
While we expect the dominant contribution to entanglement entropy to come from states localized near the boundary, there should be a finite contribution from correlations at a distance of more than one lattice spacing.

It is tempting to speculate on the relation between the local terms in the entanglement entropy and the contact term found in Ref.~\cite{Kabat1995}, since both appear to be unique to gauge theories and both are associated with observables localized on the boundary.
However our result cannot explain the negative coefficient associated with the contact term, as the local part of the entanglement entropy is manifestly positive.
There remains the intriguing possibility that the calculation of Ref.~\cite{Kabat1995} corresponds to a different definition of the entropy associated with a region of space than the one considered here, a possibility we leave for future work.

\section*{Acknowledgments}
The author acknowledges many helpful discussions with Ted Jacobson.
This research was supported in part by the Foundational Questions Institute (Grant No. RFP20816) and by NSERC.

\bibliographystyle{utphys}
\bibliography{lattice}

\end{document}